# Dualheap Sort Algorithm:
# An Inherently Parallel Generalization of Heapsort

Greg Sepesi
Eduneer, LLC
sepesi@eduneer.com
2007 June 20

**ABSTRACT**

A generalization of the heapsort algorithm is proposed. At the expense of about 50% more comparison and move operations for typical cases, the dualheap sort algorithm offers several advantages over heapsort: improved cache performance, better performance if the input happens to be already sorted, and easier parallel implementations.

**1. ALGORITHM OVERVIEW**

A heap is an array with elements regarded as nodes in a complete binary tree, where node j is the parent of nodes 2j and 2j+1, and where the value of each parent node is superior to the values of its children. This superiority of all the parent nodes is commonly called the heap condition.

Heapsort is an elegant and popular sorting algorithm by J.W.J. Williams [1] that constructs a heap and then repeatedly exchanges the values at the heap's root and its last node before decrementing the heap size and reestablishing the heap condition by performing DownHeap() on the new root value. The best and worst case running time of heapsort is Nlog(N). [2]

Heapsort is one of the most efficient sorting algorithms in terms of its low number of comparison and move operations. However, a few of heapsort's other characteristics are less efficient:

- Heapsort does not take advantage of already sorted input, performing about the same number of comparison and move operations whether or not the input is already sorted.
- Heapsort typically causes more cache misses than other sorting algorithms because each DownHeap operation typically promotes a value all the way from the root to the bottom of the heap, accessing heap values over a wide range of memory addresses.
- Aside from the initial heap construction, heapsort is not particularly well suited as an algorithm for parallel processors.

The proposed dualheap sort algorithm employs the dualheap selection algorithm [3] to recursively partition subheaps in half as shown in Figure 1-1. When the subheaps become too small to partition any more, the array is sorted. In the figure, the downward triangles represent the partitions with larger values, the upward triangles represent the partitions with smaller values, and the arrow represents the direction of increasing values and increasing heap indices, indicating that the subheaps with smaller values employ negative indices.

At the expense of about 50% more comparison and move operations for the typical case, the dualheap sort algorithm offers significant improvements to heapsort's small but important set of inefficiencies:

- When the input is already sorted, dualheap sort performs zero move operations and just NlogN comparison operations.
- The dualheap sort has better cache performance because each level of partitioning decreases the maximum promotion distance. Also dualheap sort's DownHeap operations typically do not promote values all the way to the bottom of the subheap (i.e., a smaller range of memory addresses are accessed).
- The dualheap sort algorithm is based upon the dualheap selection algorithm, which is well suited for parallel processing.

Dualheap sort also offers a significant improvement over quicksort: its stack is well-behaved and predictable (i.e., 2 lg(N) stack frames) due to the partitions being based upon precise median positions rather than median value estimates.

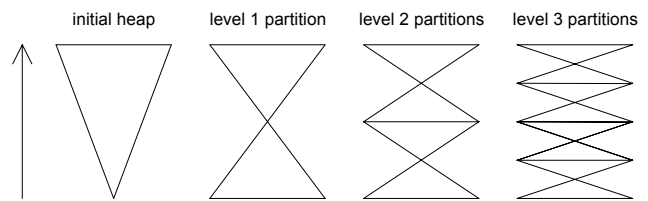

**Figure 1-1. Dualheap Sort Recursive Partitioning**

## 2. ALGORITHM DETAILS

The dualheap sort algorithm is based upon the dualheap selection algorithm, which constructs two opposing subheaps and then exchanges values between them until the larger values are in one subheap and the smaller values are in the other. Conceptually, heapsort is a special case of dualheap sort in which the size of the small subheap is one.

```
Heapsort() {
   int i, m, tmp;

   for (i=N/2; i>0; i--)
      DownHeap(i);
   for (m=N; m>1; m--, DownHeap(1)) {
      tmp=a[1]; a[1]=a[m]; a[m]=tmp;
   }
}
```

**Listing 2-1.  Heapsort**

Although heapsort is one of the most efficient sorting algorithms in terms of the low number of comparison and move operations it performs, a relatively simple modification to exchange two elements in each execution of the Heapsort loop reduces the number of comparison and move operations by another N / 2.

```
HeapsortModified() {
   int i, m, tmp;

   for (i=N/2; i>0; i--)
      DownHeap(i);
   for (m=N;
      m>3;
      m-=2, DownHeap(i), DownHeap(1))
   {
      tmp=a[1]; a[1]=a[m]; a[m]=tmp;
      i = a[2] > a[3] ? 2 : 3;
      tmp=a[i]; a[i]=a[m-1]; a[m-1]=tmp;
   }

   if (m==3) {
      a[m] = a[1];
      i = a[2] > a[3] ? 2 : 3;
      a[m-1] = a[i];
      a[m-2] = a[i^1];
   }
   else {
      a[m] = a[1];
      a[m-1] = a[2];
   }
}
```

**Listing 2-2.  Heapsort – Two Exchanges per Loop**

The modification results in N / 2 fewer move operations because every other DownHeap operation saves a move by starting one level lower in the heap. And the modification results in N / 2 fewer comparison operations because every other DownHeap, by starting one level lower in the heap, saves two comparisons, one of which is negated by the extra comparison in the modified Heapsort's for loop. This modified heapsort can be thought of as another special case of dualheap sort in which the size of the small subheap is two.

It is possible to increase the size of the small subheap to more than two, but Heapsort's loop would quickly become very complex in order to ensure that all the values moved to the small subheap are moved to their final sort position.

The dualheap sort algorithm does not constrain exchanges to only those that move values to their final sort position. Instead, it constrains moves only enough for values to be in the correct partition. As the partitions get smaller, the values eventually arrive at their sort positions. In essence, dualheap sort performs a binary search for each value as they navigate their ways to their final sort positions.

Listings 2-3, 2-4, 2-5 and 2-6 show DualheapSort, PartitionHeap, TreeSwap and DownHeap, respectively. The 'S' and 'L' postfixes denote the small valued subheap the large valued subheap, respectively. For example, pL is a pointer to a large valued subheap, and DownHeapL is a DownHeap operation on it.

```
DualheapSort(int *ph, int n) {
   int i, tmp;

   /* Construct initial heap. */
   pL = ph;
   nL = n;
   for (i=nL/2; i>0; i--)
      DownHeapL(i);

   /* Set aside two sorted elements. */
   if (pL[2] > pL[3]) {
      tmp=pL[2]; pL[2]=pL[3]; pL[3]=tmp;
   }

   /* Recursively partition. */
   if (n > 3)
      PartitionHeap(pL+2, n-2);
}
```

**Listing 2-3.  Dualeap Sort**

```
PartitionHeap(int *ph, int n) {
   int i, j, tmp;
   int nS, nL;

   /* Construct small subheap. */
   nS = (n / 2) & ~1;
   pS = ph + nS + 1;
   for (i=nS/2; i>0; i--)
      DownHeapS(i);

   /* Construct large subheap. */
   nL = n - nS;
   pL = ph + nS;
   for (i=nL/2; i>0; i--)
      DownHeapL(i);

   /* Exchange (TreeSwap phase). */
   while (pS[-1] > pL[1])
      TreeSwap(1, 1);

   /* Set aside sorted elements. */
   if (pS[-2] < pS[-3]) {
    tmp=pS[-2];pS[-2]=pS[-3];pS[-3]=tmp;
   }
   if (pL[2] > pL[3]) {
    tmp=pL[2]; pL[2]=pL[3]; pL[3]=tmp;
   }

   /* Continue partitioning. */
   if (nS > 3)
      PartitionHeap(ph, nS-2);
   if (nL > 3)
      PartitionHeap(ph+nS+2, nL-2);
}
```

**Listing 2-4. Partition Heap**

```
TreeSwap(int kS, int kL) {
   int jS, jL, tmp;

   jS = 2 * kS;
   jL = 2 * kL;
   if ((jS<=nS) && (jL<=nL)) {
      jS += (pS[-jS-1] > pS[-jS])?1:0;
      jL += (pL[jL+1] < pL[jL])?1:0;
      if (pS[-jS] > pL[jL]) {
         TreeSwap(jS, jL);
         if (pS[-(jS^1)] > pL[jL^1])
            TreeSwap(jS^1, jL^1);
      }
   }
   tmp = pS[-kS];
   pS[-kS] = pL[kL];
   pL[kL] = tmp;
   DownHeapS(kS);
   DownHeapL(kL);
}
```

**Listing 2-5. Tree Swap**

```
void DownHeapS(int k) {
   int j, v;

   if (k <= nS/2) {
      v = pS[-k];
      for (j=2*k,
         j+=(pS[-j-1] > pS[-j])?1:0;
         pS[-j] > v;
         k=j, j=2*k,
         j+=(pS[-j-1] > pS[-j])?1:0)
      {
         pS[-k] = pS[-j];
         if (j > nS/2) {
            k = j;
            break;
         }
      }
      pS[-k] = v;
   }
}

void DownHeapL(int k) {
   int j, v;

   if (k <= nL/2) {
      v = pL[k];
      for (j=2*k,
         j+=(pL[j+1] < pL[j])?1:0;
         pL[j] < v;
         k=j, j=2*k,
         j+=(pL[j+1] < pL[j])?1:0)
      {
         pL[k] = pL[j];
         if (j > nL/2) {
            k = j;
            break;
         }
      }
      pL[k] = v;
   }
}
```

**Listing 2-6. DownHeap**

As shown in Figure 1-1, each level of recursion partitions the subheaps in half. Therefore there are lg(N) levels of partitions and the running time of dualheap sort is lg(N) times the running time of the dualheap selection algorithm that performs the partitioning. Although the heap construction phases of the partition are known to be linear time, the analysis of the TreeSwap phase is still an open problem. Based upon empirical tests that show a linear run time behavior for typical cases and based upon the observation that the worst case running time of heap algorithms is generally not much different than their typical running time [4], the running time of the TreeSwap phase appears to be linear.

## 3. EMPIRICAL COMPARISON

Figures 3-1, 3-2 and 3-3 illustrate the typical performance of the dualheap sort, the heapsort, and the modified heapsort algorithms respectively, by showing the number of comparison and move operations performed by 32,000 test cases that sort uniformly distributed and pseudo-randomly generated input. Comparing the figures, dualheap sort typically performs about 50% more comparison and move operations than heapsort.

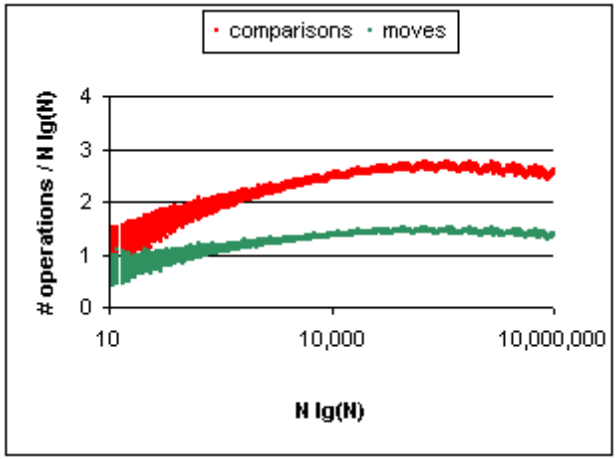

**Figure 3-1. Operations by Dualheap Sort**

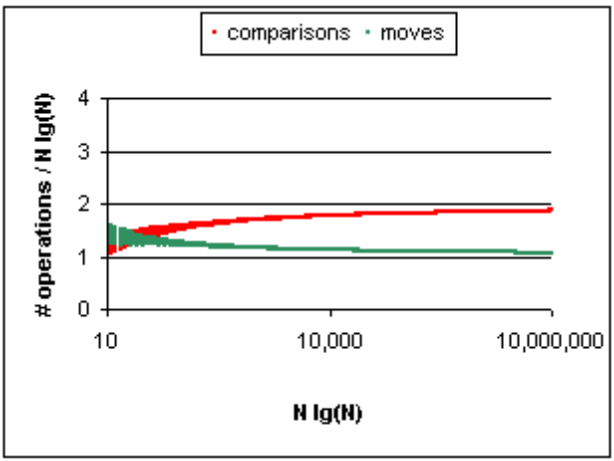

**Figure 3-2. Operations by Heapsort**

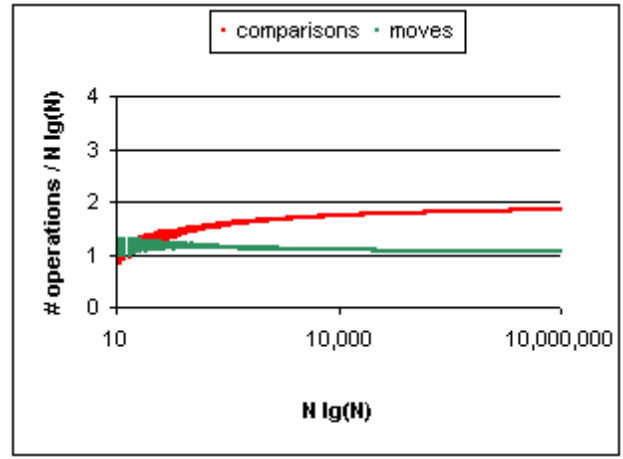

**Figure 3-3. Operations by Modified Heapsort**

## 4. CONCLUSION

At the expense of typically 50% more comparison and move operations, the dualheap sort algorithm offers improvements over heapsort in cache performance, handling of already sorted input, and parallel implementations. And over quicksort, it offers improvements in stack reliability. However, the dualheap sort algorithm's worst case analysis remains an open problem. The source code, test cases, spreadsheets and graphs presented in this paper may be a convenient starting point.[5] If you achieve even partial success in analyzing the algorithm, "the author will be pleased to know the details as soon as possible." Ending with that quote from D.E. Knuth [6] is fitting, as the proposed sorting algorithm would likely not exist without his encyclopedic coverage of the topic.

## 5. REFERENCES

[1] J.W.J.Williams, *Algorithm 232 - Heapsort*, Communications of the ACM, 7 (12), p.347-348, 1964.
[2] R.Sedgewick, *Algorithms*, ISBN: 0-201-06673-4, p.157, 1988.
[3] G.Sepesi, *Dualheap Selection Algorithm Source Code, Test Cases, Spreadsheets and Graphs*, http://eduneer.com/pub/dh.zip, 2007.
[4] D.E.Knuth, *The Art of Computer Programming – Volume 3*, chapter 5.2.3 Sorting by Selection, ISBN: 0-201-03803-X, p.149, 1973.
[5] G.Sepesi, *Dualheap Selection Algorithm Source Code, Test Cases, Spreadsheets and Graphs*, http://eduneer.com/pub/dhs.zip, 2007.
[6] D.E.Knuth, *The Art of Computer Programming – Volume 3*, Notes on Exercisesc, ISBN: 0-201-03803-X, p.ix, 1973.